\documentclass[twocolumn]{openjournal}

\usepackage{natbib}
\usepackage{graphicx,amsmath,amssymb,amstext}
\usepackage{amsbsy,amsfonts,amsthm,color}
\usepackage[colorlinks,linkcolor=blue,citecolor=blue,urlcolor=blue ]{hyperref}
\usepackage{float}
\usepackage{subcaption}
\usepackage{enumerate}
\usepackage{multirow}

\begin{document}

\title{A Deep Multimodal Multi--Head Neural Network for Joint Estimation of Stellar Age, Lifetime, and  Evolutionary Stage} 

\author{Jing Rou Puah}
\affiliation{Faculty of AI \& Robotics, Raffles University, Johor Bahru, Malaysia}
\email{$^*$email: puahjingrou@raffles.university}

\author{Sasa Arsovski}
\affiliation{Faculty of AI \& Robotics, Raffles University, Johor Bahru, Malaysia}

\begin{abstract}
\noindent
The accurate estimation of stellar parameters: stellar age, lifespan, and evolutionary stage remain a fundamental challenge in astrophysics. This paper introduces novel deep learning architecture that combines multimodal spectroscopic and photometric data from the Sloan Digital Sky Survey (SDSS) to estimate stellar parameters. We utilise multimodal data from SDSS Data Release 17, creating a 29-dimensional feature set from quality-filtered, extinction-corrected photometry and applying standardised preprocessing to all spectra. The corresponding training labels are derived from MIST v1.2 isochrones: logarithmic age and lifetime are calculated from atmospheric parameters, while evolutionary stage is binned into five classes (Hot, Medium, Cool, Subgiant, Red Giant) based on the star's position in the $T_{\mathrm{eff}}$--$\log(g)$ diagram. We conduct multi-phase evaluation of eight different deep learning architectures to assess their performance and generalisation capability. In the Phase I, we explore the relationship between model architectures and data balancing strategies. Phase II focuses on the systematic tuning of architectural complexity (depth and width), while Phase III optimises the composition of the multi-task loss function to achieve the best trade-off between regression and classification performance. Following multi-phase evaluation, the final model architecture was selected, a novel hybrid deep learning architecture for stellar parameter estimation. The architecture comprises two main branches: a Multi-Layer Perceptron (MLP) for numerical stellar data analysis (e.g., temperature) and a Convolutional Neural Network (CNN) combined with a Vision Transformer (ViT) for stellar spectrogram analysis. The output layer includes three output heads: one for stellar age prediction, one for stellar lifetime prediction, and one for stellar evolutionary stage classification. The exploration of architectural complexity and the tuning of multi-task objective weights revealed that the selected final architecture achieved the best balance between mathematical precision (e.g., an Age RMSE of 0.093 in ($\log(\mathrm{yrs})$) space) and physical realism. To evaluate robustness, we employed Monte Carlo Dropout and confirmed that the model produces well-calibrated uncertainties that accurately reflect prediction errors. These robust uncertainties enable meaningful astrophysical interpretation and set a new benchmark for multimodal stellar parameter estimation.
\end{abstract}

\keywords{Deep Learning, Multimodal Learning, Astro-informatics, Stellar Parameters, Multi-Task Learning}

\section{Introduction}
The accurate estimation of a star's fundamental parameters: age, total lifetime, and evolutionary stage is a foundation of modern astrophysics. These parameters form the chronological framework of the universe, enabling research that ranges from reconstructing the formation history of the Milky Way to assessing the habitability of exoplanetary systems \citep{maeder2008physics}. 

However, despite their importance, deriving these parameters is extremely difficult. As noted by \citet{soderblom2010ages}, stellar age is “the most important single parameter of a star, yet it is one of the most difficult to determine”. The primary challenge lies in parameter degeneracy, where different combinations of intrinsic properties (e.g., age, mass, metallicity) and extrinsic factors (e.g., interstellar dust) produce nearly indistinguishable observational signatures \citep{ksoll2020stellar}. This is particularly problematic for the vast population of field stars, which constitute the majority of observable samples but lack the uniform age distribution that characterises star clusters. This degeneracy leads to three critical limitations in current methodologies:

\begin{enumerate}
    \item Uncertainty from Single-Modality Data: Relying on either photometry or spectroscopy alone is often insufficient to break the degeneracy, leading to significant uncertainties in age and lifetime estimation \citep{gai2024simultaneous}.
    
    \item Ambiguity in Stage Classification: Distinguishing between subtle, contiguous evolutionary stages (e.g., late main-sequence vs. early subgiant) requires sophisticated feature extraction that is often treated as a separate problem from age regression.
    
    \item Lack of a Unified Framework: Age, lifetime are inherently linked to its evolutionary stage, yet they are typically addressed as separate tasks, a disjointed approach that fails to leverage their mutual constraints.
\end{enumerate}

The advent of large-scale surveys like the Sloan Digital Sky Survey (SDSS) \citep{stoughton2002sloan} provides a unique opportunity to address these challenges. By providing co-located, multimodal data --- broadband photometry and detailed spectroscopy --- for millions of objects, SDSS enables a data-driven approach to simultaneously resolve these parameters. While machine learning methods have proven effective in this domain \citep{gai2024simultaneous, li2025machine}, this work presents the first systematic comparison of modern, deep multimodal architectures for this specific multi-task problem.

We explore the relationship between model architectures, data balancing strategies, and the composition of the multi-task objective function, leading to the selection of a final model based on both mathematical precision and its adherence to physical realism --- a critical criterion for scientific applications.

The key contributions of this work are as follows:

\begin{itemize}
    \item It presents a systematic, comparative analysis of eight deep learning architectures, providing empirical evidence on the trade-offs between classic (ResNet-based) and modern (Transformer-based) models for this task.
    
    \item It demonstrates two critical findings: first, that a data-level under sampling strategy can unexpectedly outperform a weighted-loss approach in this domain; and second, that a fundamental trade-off exists between mathematical precision and physical realism in the final multi-task model.
    
    \item It delivers a rigorously validated, high-fidelity regression model with quantified uncertainty, which achieves a state-of-the-art performance (Age RMSE of 0.093 in $\log(\mathrm{yrs})$ space) while ensuring all predictions are scientifically valid, providing a reusable framework for future studies.
\end{itemize}

\section{Related Work}
Our research is positioned at the intersection of three key domains: stellar astrophysics, deep learning for scientific applications, and multimodal, multi-task learning. This section briefly reviews the foundational work in each area to establish the context and identify the specific research gap our study addresses.

\subsection{Stellar Parameter Estimation}
The estimation of stellar age, lifetime, and evolutionary stage has long been a central pursuit in astrophysics. The gold-standard method --- isochrone fitting on colour-magnitude diagrams --- provides reliable ages for coeval stellar populations in clusters \citep{bell2001stellar}. However, its application to isolated field stars --- which constitute the vast majority of galactic populations --- is plagued by significant uncertainties arising from the inherent degeneracy between stellar parameters \citep{soderblom2010ages}. While alternative methods like asteroseismology \citep{chaplin2013asteroseismology} and gyrochronology \citep{barnes2008gyrochronology} offer high precision for specific types of stars, they are not universally applicable, leaving a critical need for robust methods that can be applied to the large, heterogeneous datasets produced by modern surveys.

\subsection{Deep Learning in Stellar Astrophysics}
The advent of deep learning has introduced powerful new tools for stellar parameter estimations. Convolutional Neural Networks (CNNs), which excel at extracting localised features \citep{lecun2015deep}, have been widely applied to one-dimensional stellar spectra \citep{wang2024deep}. Architectures based on ResNet, for example, have become a benchmark for spectral analysis \citep{he2016deep}. More recently, the Transformer architecture, with its self-attention mechanism, has shown significant promise in modelling the long-range dependencies within spectra, which is crucial for separating the interdependent influences of multiple physical parameters \citep{vaswani2017attention}. However, most existing studies have applied these architectures either to single-modality data (spectroscopy or photometry alone) or to single-task objectives. A systematic, direct comparison of the trade-offs between classic (ResNet-based) and modern (Transformer-based) approaches within a unified, multimodal framework remains a key research gap.

\subsection{Multimodal and Multi-Task Learning}
Fusing complementary data sources through multimodal learning is a powerful strategy to mitigate the degeneracy problem \citep{gai2024simultaneous}. Photometry provides broadband colour and luminosity information, which constrains temperature and distance, while spectroscopy reveals detailed atmospheric physics and chemical composition. While this approach has demonstrated effectiveness in related fields such as galaxy studies, its application to stellar parameter estimation is still maturing.

Furthermore, stellar age, lifetime, and stage are not independent variables; they are intrinsically correlated through the laws of stellar evolution. A multi-task learning (MTL) framework allows the model to achieve better generalisation by utilising the shared information between related task --- a concept systematically established by \citet{caruana1997multitask} --- to learn a more coherent and physically-grounded latent representation. However, this introduces its own challenges. Prior studies have often overlooked two critical, practical aspects:

\begin{enumerate}
    \item The profound impact of data balancing strategies (e.g., data-level undersampling vs. algorithm-level weighting) on the performance of both classification and regression tasks in a real-world, severely imbalanced astronomical context.
    
    \item The necessity of a completely evaluation criterion that extends beyond standard mathematical metrics (e.g., RMSE) to include the physical realism of predictions, a crucial consideration for any scientific machine learning model.
\end{enumerate}

This work is designed to directly address these gaps by conducting a systematic, end-to-end comparison of multimodal architectures and training strategies, with a unique focus on the relationship between data balancing, mathematical precision, and scientific validity.

\section{Methodology}
\begin{figure*}[t]
    \centering
    \includegraphics[width=\textwidth]{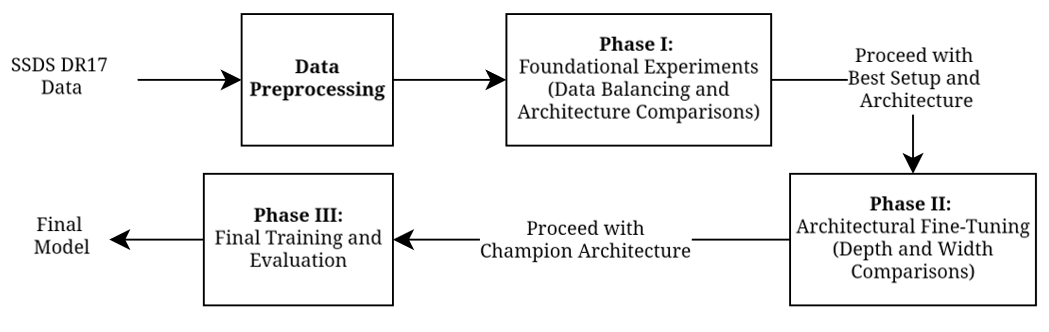}
    \caption{The overall three-phase research workflow.}
    \label{fig:workflow}
\end{figure*}

The estimation of stellar age, lifetime, and methodology of this research follows a structured, three-phase workflow designed to systematically identify the optimal model and training strategy for multi-task stellar parameter estimation. The overall workflow is illustrated in Fig.~\ref{fig:workflow}.

\subsection{Data and Data Acquisition}
For data consistency, the photometric and spectroscopic data for our multimodal dataset were sourced exclusively from the Sloan Digital Sky Survey Data Release 17 (SDSS DR17) \citep{stoughton2002sloan}. The initial dataset was constructed by querying the SDSS CasJobs SQL server, joining the fundamental \texttt{SpecObj} (spectroscopic metadata), \texttt{sppParams} (stellar parameters from the SSPP pipeline), and \texttt{PhotoObj} (photometric measurements) tables. To ensure the reliability of the final catalogue, a set of strict quality criteria was applied, including the selection of confirmed stars (\texttt{class = ‘STAR’}), high-precision redshift measurements (\texttt{zErr < 0.0002}), a high median signal-to-noise ratio (\texttt{snMedian > 20}), and reliable atmospheric parameter estimations with low uncertainties.

\subsection{Data Preprocessing Pipeline}
To prepare the raw observational data for our deep learning models, we implemented a standardised pipeline to create a clean, uniform, and feature-rich dataset. This process involved four key stages.

\subsubsection{Data Filtering} 
A multi-stage quality assurance protocol was applied. This included enforcing photometric precision (\texttt{psfMagErr\_* < 0.1} mag), removing entries with non-physical negative extinction values, and cleaning the parameter space to ensure all stars fell within the valid range ($T_{\mathrm{eff}} \in \mathrm{K}$) of the theoretical models used for labelling.

\subsubsection{Label Generation and Engineering} 
As stellar age and lifetime are not directly observable, ground-truth labels were generated using the MESA Isochrones and Stellar Tracks (MIST) v1.2 models \citep{choi2016mesa}. For each star, its high-fidelity atmospheric parameters ($T_{\mathrm{eff}}$, $\log(g)$, $\mathrm{[Fe/H]}$) were used as inputs to query the MIST grid, yielding estimates for age and initial mass. The lifetime ($\tau$) was then derived from the initial mass ($M$) via the standard mass--lifetime relation ($\tau \propto M^{-2.5}$) \citep{maeder2008physics}. To handle their wide dynamic range, both age and lifetime were converted to a logarithmic scale ($\log(\mathrm{age}/\mathrm{yrs}$) and $\log(\mathrm{lifetime}/\mathrm{yrs})$), which is a standard practice. The categorical stage label was engineered by applying thresholds in the $T_{\mathrm{eff}}$--$\log(g)$ space to define five distinct evolutionary classes: Hot, Medium, Cool, Subgiant, and Red Giant.

\subsubsection{Photometric Processing} 
To better represent the intrinsic physical properties, we derived a 29-dimensional feature vector from the raw photometry. This involved:
    
    \begin{enumerate}[(i)]
        \item Extinction Correction, where interstellar dust effects were removed by subtracting extinction values from the raw PSF magnitudes to produce \texttt{corrected\_psfMag\_* } features; and
        
        \item Colour Index Calculation, where four standard colour indices (e.g., $u-g$, $g-r$) were computed from these corrected magnitudes.
    \end{enumerate}

\subsubsection{Spectroscopic Processing} 
The downloaded FITS files underwent a standardised, four-step pipeline:

    \begin{enumerate}[(i)]
        \item Redshift Correction to the rest frame;
        
        \item Resampling via linear interpolation onto a common wavelength grid (3,600~\AA\ to 10,400~\AA, 1.0~\AA\ step), creating a uniform 6800-point vector; 
        
        \item Denoising using a Savitzky--Golay filter \citep{savitzky1964smoothing}; and
        
        \item Continuum Normalisation by dividing each spectrum by its median flux. A final quality control step removed any remaining "flat" spectra (standard deviation $< 0.01$).
    \end{enumerate}

\subsection{Experimental Design}
Our multi-phase experimental framework, detailed below, was designed to systematically navigate the challenges of data imbalance and architectural complexity.

\subsubsection{Phase I: Foundational Experiments}
This phase aimed to understand the impact of class imbalance. We created three experimental setups to analyse data imbalance: Experiment A (50k imbalanced baseline), Experiment B (23k balanced via under sampling), and Experiment C (50k imbalanced with a weighted loss). Eight architectures was evaluated on these setups to identify the most effective balancing strategy and shortlist finalist models, with their configurations shown in Table~\ref{tab:table-i}.

\begin{table}[t]
\centering
\caption{Summary of model architectures.}
\begin{tabular}{|clccc|}
\hline
\multicolumn{1}{|c|}{ID} & \multicolumn{1}{c|}{\begin{tabular}[c]{@{}c@{}}Model\\ Name\end{tabular}}           & \multicolumn{1}{c|}{\begin{tabular}[c]{@{}c@{}}Photometric\\ Branch\end{tabular}} & \multicolumn{1}{c|}{\begin{tabular}[c]{@{}c@{}}Spectroscopic\\ Branch\end{tabular}} & \begin{tabular}[c]{@{}c@{}}Attention\\ Type\end{tabular}      \\ \hline
\multicolumn{5}{|c|}{Single-Modality Baselines}                                                                                                                                                                                                                                                                                                       \\ \hline
\multicolumn{1}{|c|}{1}  & \multicolumn{1}{l|}{MLP-Only}                                                       & \multicolumn{1}{c|}{MLP}                                                         & \multicolumn{1}{c|}{None}                                                          & N/A                                                           \\ \hline
\multicolumn{1}{|c|}{2}  & \multicolumn{1}{l|}{ResNet-Only}                                                    & \multicolumn{1}{c|}{None}                                                        & \multicolumn{1}{c|}{ResNet}                                                        & N/A                                                           \\ \hline
\multicolumn{1}{|c|}{3}  & \multicolumn{1}{l|}{\begin{tabular}[c]{@{}l@{}}Spectra-\\ Transformer\end{tabular}} & \multicolumn{1}{c|}{None}                                                        & \multicolumn{1}{c|}{Transformer}                                                   & \begin{tabular}[c]{@{}c@{}}Standard\\ Multi-Head\end{tabular} \\ \hline
\multicolumn{1}{|c|}{4}  & \multicolumn{1}{l|}{\begin{tabular}[c]{@{}l@{}}Spectra-\\ Hadamard\end{tabular}}    & \multicolumn{1}{c|}{None}                                                        & \multicolumn{1}{c|}{Transformer}                                                   & \begin{tabular}[c]{@{}c@{}}Hadamard\\ Gated\end{tabular}      \\ \hline
\multicolumn{5}{|c|}{Multimodal Models}                                                                                                                                                                                                                                                                                                              \\ \hline
\multicolumn{1}{|c|}{5}  & \multicolumn{1}{l|}{MLP+ResNet}                                                     & \multicolumn{1}{c|}{MLP}                                                         & \multicolumn{1}{c|}{ResNet}                                                        & N/A                                                           \\ \hline
\multicolumn{1}{|c|}{6}  & \multicolumn{1}{l|}{\begin{tabular}[c]{@{}l@{}}MLP+CNN+\\ Transformer\end{tabular}} & \multicolumn{1}{c|}{MLP}                                                         & \multicolumn{1}{c|}{\begin{tabular}[c]{@{}c@{}}CNN+\\ Transformer\end{tabular}}    & \begin{tabular}[c]{@{}c@{}}Standard\\ Multi-Head\end{tabular} \\ \hline
\multicolumn{1}{|c|}{7}  & \multicolumn{1}{l|}{\begin{tabular}[c]{@{}l@{}}MLP+CNN+\\ Hadamard\end{tabular}}    & \multicolumn{1}{c|}{MLP}                                                         & \multicolumn{1}{c|}{\begin{tabular}[c]{@{}c@{}}CNN+\\ Transformer\end{tabular}}    & \begin{tabular}[c]{@{}c@{}}Hadamard\\ Gated\end{tabular}      \\ \hline
\multicolumn{1}{|c|}{8}  & \multicolumn{1}{l|}{\begin{tabular}[c]{@{}l@{}}MLP+\\ Transformer\end{tabular}}     & \multicolumn{1}{c|}{MLP}                                                         & \multicolumn{1}{c|}{Transformer}                                                   & \begin{tabular}[c]{@{}c@{}}Standard\\ Multi-Head\end{tabular} \\ \hline
\end{tabular}
\label{tab:table-i}
\end{table}

\subsubsection{Phase II: Architectural Fine-Tuning}
This phase focused on optimising the complexity of the most promising architecture from Phase~I. Conducted on the balanced Experiment B dataset, this process involved a coordinate ascent methodology, sequentially tuning the depth and width of each key model component (Spectroscopic, Photometric, and Fusion). The full hyperparameter space explored in this phase, designed to explore the model's response to changes in feature abstraction depth and representational capacity, is detailed in Table~\ref{tab:table-ii}.

\begin{table}[t]
\centering
\caption{Hyperparameter space for Phase II.}
\begin{tabular}{|c|l|l|}
\hline
Component                                                                       & \multicolumn{1}{c|}{Parameter}                                             & \multicolumn{1}{c|}{\begin{tabular}[c]{@{}c@{}}Values\\ Explored\end{tabular}} \\ \hline
\multirow{2}{*}{\begin{tabular}[c]{@{}c@{}}Spectroscopic\\ Branch\end{tabular}} & \begin{tabular}[c]{@{}l@{}}Depth\\ (\texttt{num\_transformer\_layers})\end{tabular} & \{2, 4, 6\}                                                                    \\ \cline{2-3} 
                                                                                & Width (\texttt{d\_model})                                                           & \{128, 256, 384\}                                                              \\ \hline
\multirow{2}{*}{\begin{tabular}[c]{@{}c@{}}Photometric\\ Branch\end{tabular}}   & Depth (\texttt{mlp\_depth})                                                         & \{2, 3, 4\}                                                                    \\ \cline{2-3} 
                                                                                & Width (\texttt{mlp\_hidden\_dim})                                                   & \{128, 256, 384\}                                                              \\ \hline
\multirow{2}{*}{\begin{tabular}[c]{@{}c@{}}Fusion\\ Module\end{tabular}}        & Depth (\texttt{fusion\_depth})                                                      & \{2, 3, 4\}                                                                    \\ \cline{2-3} 
                                                                                & Width (\texttt{fusion\_hidden\_dim})                                                & \{256, 512, 768\}                                                              \\ \hline
\end{tabular}
\label{tab:table-ii}
\end{table}

\subsubsection{Phase III: Final Training and Evaluation}
The final phase focused on training the best-performing architecture on the full 166,625-star dataset. This involved an exploratory, iterative process to identify a stable multi-task training strategy, including an investigation into the optimal composition of the loss function and the trade-off between mathematical precision and scientific validity.

\subsection{Model Architectures}
We implemented eight foundational architectures, grouped into three categories and summarised in Table~\ref{tab:table-i}. The descriptions below pertain to their baseline configurations.

\subsubsection{Single-Modality Baselines (Model 1, 2)}
An MLP-only model to establish the performance ceiling of photometry, and a ResNet-based \citep{he2016deep} CNN to serve as a benchmark for classic spectral analysis.

\subsubsection{Spectral-Only Transformers (Model 3, 4)}
These models consist only of the Transformer spectroscopy branch, serving a dual purpose. First, they provide a direct modern benchmark against the classic ResNet-Only baseline. Second, they enable a clean A/B test between the standard Scaled Dot-Product Attention \citep{vaswani2017attention} and our custom Hadamard Gated Attention. Inspired by Gated Linear Units (GLU) \citep{dauphin2017language}, our custom mechanism uses a computationally lighter element-wise gating function that computes the output as $y = \text{sigmoid}(QK) \odot V$.

\subsubsection{Multimodal Models (Model 5, 6, 7, 8)} 
\begin{figure*}[t]
    \centering
    \includegraphics[width=\textwidth]{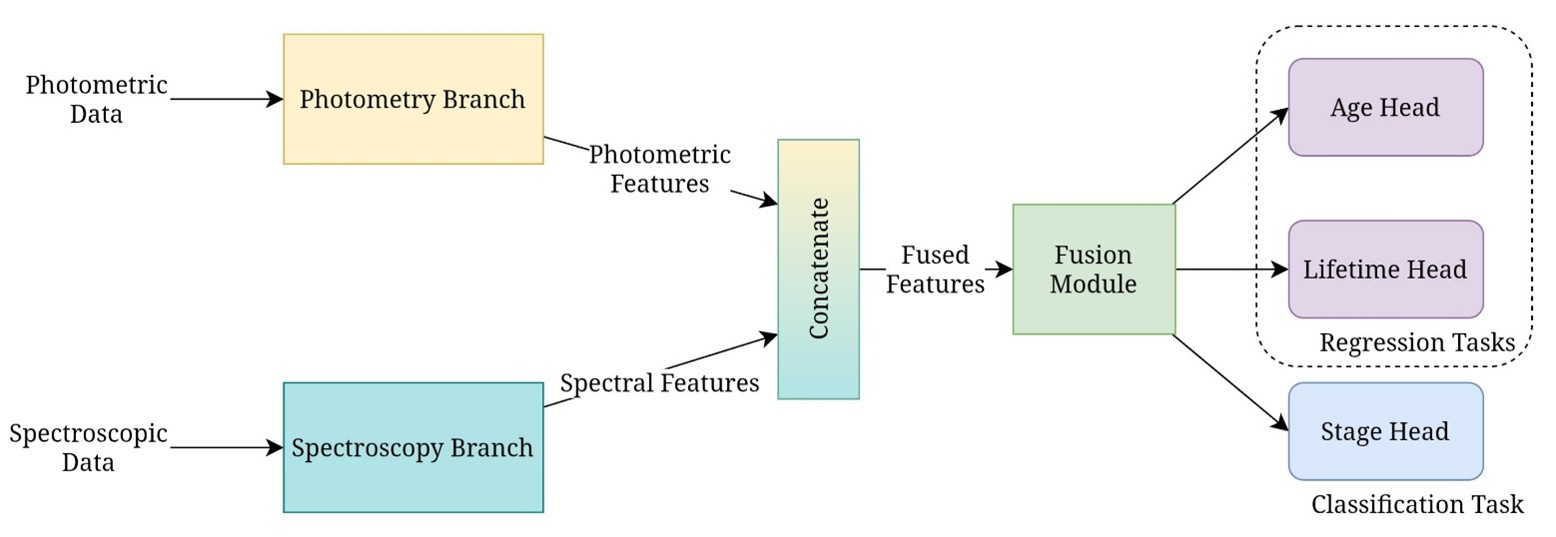}
    \caption{General multimodal framework diagram.}
    \label{fig:multimodal}
\end{figure*}

All our multimodal models follow the general framework shown in Fig.~\ref{fig:multimodal}, combining an MLP for photometry with a specialised branch for spectroscopy. These models form the core of our investigation, designed to test different strategies for fusing photometric and spectroscopic data. Key comparisons include a classic vs. modern showdown (MLP+ResNet vs. MLP+CNN+Transformer) and an ablation study on the value of the CNN pre-processor (MLP+Transformer vs. MLP+CNN+Transformer).

\subsection{Training and Evaluation Setup}
A unified setup was used across all phases to ensure fair comparison.

\begin{enumerate}
    \item Training: Models were trained using the AdamW optimizer \citep{loshchilov2017decoupled} with a baseline learning rate of $1\times10^{-5}$ and a weight decay of $1\times10^{-2}$. The learning rate was managed by a \texttt{ReduceLROnPlateau} scheduler. We adopted a multi-task learning approach, with a total loss comprising a weighted sum of mean-squared-error (MSE) losses for regression and a cross-entropy loss for classification. For experiments on imbalanced data, the classification loss was weighted using Scikit-learn’s \texttt{compute\_class\_weight} function \citep{geron2022hands}. A checkpointing system was implemented to allow for recovery from interruptions.

    \item Evaluation: Performance was quantified using macro F1-score for classification and RMSE/MAE for regression in both $\log(\mathrm{yrs})$ and ($\mathrm{Gyr}$) units. A critical component of our evaluation was the assessment of physical realism. For the final model, we implemented Monte Carlo dropout \citep{gal2016dropout} to quantify prediction uncertainty.
\end{enumerate}

\section{Results and Discussion}
This chapter presents a detailed analysis of the results from our three-phase experimental framework. We first establish the optimal data handling strategy, then detail the architectural fine-tuning process to identify a best-performing model, and finally, present the performance of this model on the full dataset, highlighting the critical trade-offs fundamental to this multi-task problem.

\subsection{Phase I: Data Balancing and Architecture Screening Results}
The initial phase of our investigation focused on addressing the foundational challenges of class imbalance and performing a broad comparative screening of the eight proposed model architectures.

Our first key finding, summarised in Table~\ref{tab:table-iii}, relates to the data handling strategy. A direct comparison revealed that the data-level under sampling approach (Experiment B) unexpectedly and significantly outperformed the algorithm-level weighted-loss approach (Experiment~C) across all tasks. For instance, F1-scores improved by a relative $33\text{--}35\%$, and regression RMSEs were reduced by $33\text{--}46\%$. This unexpectedly favourable result suggests that for this dataset, the negative impact of the majority class dominating the feature space was more detrimental than the information loss from under sampling. This established Experiment~B as the most effective and unbiased environment for a fair comparison of model architectures.

\begin{table}[t]
\centering
\caption{Performance comparison between Experiment B and C.}
\begin{tabular}{|c|l|c|c|}
\hline
Model                    & \multicolumn{1}{c|}{Metrics}                & \begin{tabular}[c]{@{}c@{}}Experiment C\\ (Weighted)\end{tabular} & \begin{tabular}[c]{@{}c@{}}Experiment B\\ (Under Sampling)\end{tabular} \\ \hline
\multirow{3}{*}{Model 6} & F1 Macro    & 0.6410                                                            & 0.8577                                                                  \\ \cline{2-4} 
                         & Age RMSE     & 0.3376                                                            & 0.1940                                                                  \\ \cline{2-4} 
                         & Lifetime RMSE & 0.3504                                                            & 0.2041                                                                  \\ \hline
\multirow{3}{*}{Model 5} & F1 Macro      & 0.6828                                                            & 0.9187                                                                  \\ \cline{2-4} 
                         & Age RMSE      & 0.3933                                                            & 0.2250                                                                  \\ \cline{2-4} 
                         & Lifetime RMSE & 0.4312                                                            & 0.2316                                                                  \\ \hline
\multirow{3}{*}{Model 1} & F1 Macro      & 0.6455                                                            & 0.8616                                                                  \\ \cline{2-4} 
                         & Age RMSE      & 0.6790                                                            & 0.4494                                                                  \\ \cline{2-4} 
                         & Lifetime RMSE & 0.6816                                                            & 0.4500                                                                  \\ \hline
\end{tabular}
\label{tab:table-iii}
\end{table}

With this optimal training environment established, we then proceeded with the initial architecture comparison. The results on Experiment~B, detailed in Table~\ref{tab:table-iv}, revealed a clear task specialisation. Single-modality models demonstrated the fundamental trade-off, with the spectra-only Transformer (Model~3) excelling at regression ($\mathrm{Age\ RMSE}=0.0901$) but failing at classification ($\mathrm{F1}=0.6213$), while the MLP-only model (Model~1) showed the opposite trend. The multimodal models, however, proved more capable of balancing the tasks. Notably, the classic MLP+ResNet (Model~5) emerged as a top-tier classifier ($\mathrm{F1}=0.9187$) with respectable regression performance. In contrast, the modern MLP+CNN+Transformer (Model~6) distinguished itself with the best regression performance among all multimodal models ($\mathrm{Age\ RMSE}=0.1940$), while still maintaining a strong classification score ($\mathrm{F1}=0.8577$). 

\begin{table}[t]
\centering
\caption{Comprehensive performance metrics on Experiment B.}
\begin{tabular}{|lccc|}
\hline
\multicolumn{1}{|c|}{Model ID and Name}      & \multicolumn{1}{c|}{\begin{tabular}[c]{@{}c@{}}F1\\ Macro\end{tabular}} & \multicolumn{1}{c|}{\begin{tabular}[c]{@{}c@{}}Age\\ RMSE\end{tabular}} & \begin{tabular}[c]{@{}c@{}}Lifetime\\ RMSE\end{tabular} \\ \hline
\multicolumn{4}{|c|}{Single-Modality Baselines}                                                                                                                                                                                                  \\ \hline
\multicolumn{1}{|l|}{1. MLP-Only}            & \multicolumn{1}{c|}{0.8616}                                             & \multicolumn{1}{c|}{0.4494}                                             & 0.4500                                                  \\ \hline
\multicolumn{1}{|l|}{2. ResNet-Only}         & \multicolumn{1}{c|}{0.7492}                                             & \multicolumn{1}{c|}{0.2428}                                             & 0.2338                                                  \\ \hline
\multicolumn{1}{|l|}{3. Spectra-Transformer} & \multicolumn{1}{c|}{0.6213}                                             & \multicolumn{1}{c|}{0.0901}                                             & 0.0892                                                  \\ \hline
\multicolumn{1}{|l|}{4. Spectra-Hadamard}    & \multicolumn{1}{c|}{0.6417}                                             & \multicolumn{1}{c|}{0.1486}                                             & 0.1576                                                  \\ \hline
\multicolumn{4}{|c|}{Multimodal Models}                                                                                                                                                                                                        \\ \hline
\multicolumn{1}{|l|}{5. MLP+ResNet}          & \multicolumn{1}{c|}{0.9187}                                             & \multicolumn{1}{c|}{0.2250}                                             & 0.2316                                                  \\ \hline
\multicolumn{1}{|c|}{6. MLP+CNN+Transformer} & \multicolumn{1}{c|}{0.8577}                                             & \multicolumn{1}{c|}{0.1940}                                             & 0.2041                                                  \\ \hline
\multicolumn{1}{|l|}{7. MLP+CNN+Hadamard}    & \multicolumn{1}{c|}{0.9294}                                             & \multicolumn{1}{c|}{0.2776}                                             & 0.2832                                                  \\ \hline
\multicolumn{1}{|l|}{8. MLP+Transformer}     & \multicolumn{1}{c|}{0.8062}                                             & \multicolumn{1}{c|}{0.2388}                                             & 0.2384                                                  \\ \hline
\end{tabular}
\label{tab:table-iv}
\end{table}

Given that the project’s primary objective is precise regression, this superior regression capability marked Model~6 as the primary finalist. Model~5 was also carried forward as a key benchmark, setting the stage for a direct classic-versus-modern architecture showdown in the fine-tuning phase.

\subsection{Phase II: Architectural Fine-Tuning Results}
We conducted a systematic fine-tuning process on the finalist architectures to determine the optimal complexity. This involved a coordinate ascent-style exploration of depth and width for each key component.

Our key findings from this phase are summarised in Table~\ref{tab:table-v}. The initial depth comparison confirmed the architectural superiority of the modern MLP+CNN+Transformer (Model~6) over the classic MLP+ResNet (Model~5). As shown in the Table V, Model~6’s 4-layer configuration achieved an Age RMSE of $0.0979$, decisively outperforming Model~5’s best result of $0.1654$.

\begin{table}[]
\centering
\caption{Performance metrics for spectroscopic branch depth tuning.}
\begin{tabular}{|lccc|}
\hline
\multicolumn{4}{|c|}{Model 6 (MLP+CNN+Transformer)}                                                                                                \\ \hline
\multicolumn{1}{|c|}{Metrics}          & \multicolumn{1}{c|}{Depth=2} & \multicolumn{1}{c|}{Depth=4} & Depth=6 \\ \hline
\multicolumn{1}{|l|}{Age RMSE ($\log(\mathrm{yrs})$)}       & \multicolumn{1}{c|}{0.1907}           & \multicolumn{1}{c|}{0.0979}           & 0.1076           \\ \hline
\multicolumn{1}{|l|}{Lifetime RMSE ($\log(\mathrm{yrs})$)}  & \multicolumn{1}{c|}{0.1912}           & \multicolumn{1}{c|}{0.1017}           & 0.1091           \\ \hline
\multicolumn{1}{|l|}{Classification F1 (macro)} & \multicolumn{1}{c|}{0.8230}           & \multicolumn{1}{c|}{0.9025}           & 0.9541           \\ \hline
\multicolumn{4}{|c|}{Model 5 (MLP+ResNet)}                                                                                                         \\ \hline
\multicolumn{1}{|c|}{Metrics}          & \multicolumn{1}{c|}{Depth=2} & \multicolumn{1}{c|}{Depth=4} & Depth=6 \\ \hline
\multicolumn{1}{|l|}{Age RMSE ($\log(\mathrm{yrs})$)}       & \multicolumn{1}{c|}{0.2518}           & \multicolumn{1}{c|}{0.1654}           & 0.1754           \\ \hline
\multicolumn{1}{|l|}{Lifetime RMSE ($\log(\mathrm{yrs})$)}  & \multicolumn{1}{c|}{0.2613}           & \multicolumn{1}{c|}{0.1680}           & 0.1816           \\ \hline
\multicolumn{1}{|l|}{Classification F1 (macro)} & \multicolumn{1}{c|}{0.8764}           & \multicolumn{1}{c|}{0.9346}           & 0.9338           \\ \hline
\end{tabular}
\label{tab:table-v}
\end{table}

Based on this, we focused exclusively on the Transformer architecture for all subsequent tuning. The next critical step, detailed in Table~\ref{tab:table-vi}, was the width ($d_{\mathrm{model}}$) exploration. This analysis revealed a crucial trade-off between mathematical precision and physical realism. While the $d_{\mathrm{model}}=128$ variant achieved a stellar F1-score ($0.9544$), it produced scientifically unrealistic age predictions ($>60~\mathrm{Gyr}$). In contrast, the $d_{\mathrm{model}}=256$ configuration produced physically valid predictions while maintaining top-tier performance.

\begin{table}[t]
\centering
\caption{Performance metrics for spectroscopic branch width tuning.}
\begin{tabular}{|llccc|}
\hline
\multicolumn{2}{|c|}{Metrics}                                                                                                                     & \multicolumn{1}{c|}{Width=128} & \multicolumn{1}{c|}{Width=256} & Width=384 \\ \hline
\multicolumn{2}{|l|}{Age RMSE (log(yrs))}                                                                                                         & \multicolumn{1}{c|}{0.0961}    & \multicolumn{1}{c|}{0.0979}    & 0.1686    \\ \hline
\multicolumn{2}{|l|}{Lifetime RMSE (log(yrs))}                                                                                                    & \multicolumn{1}{c|}{0.0993}    & \multicolumn{1}{c|}{0.1017}    & 0.1726    \\ \hline
\multicolumn{2}{|l|}{Classification F1 (macro)}                                                                                                   & \multicolumn{1}{c|}{0.9544}    & \multicolumn{1}{c|}{0.9025}    & 0.9253    \\ \hline
\multicolumn{5}{|c|}{True Age Range (Gyr): 2.842 $\sim$6.107; Mean: 5.022}                                                                                                                                             \\ \hline
\multicolumn{1}{|l|}{\multirow{3}{*}{\begin{tabular}[c]{@{}l@{}}Predicted\\ Age Range\\ (Gyr)\end{tabular}}} & \multicolumn{1}{l|}{Min}  & \multicolumn{1}{c|}{0.737}     & \multicolumn{1}{c|}{1.105}     & 2.343     \\ \cline{2-5} 
\multicolumn{1}{|l|}{}                                                                                       & \multicolumn{1}{l|}{Max}  & \multicolumn{1}{c|}{68.704}    & \multicolumn{1}{c|}{14.870}    & 267.134   \\ \cline{2-5} 
\multicolumn{1}{|l|}{}                                                                                       & \multicolumn{1}{l|}{Mean} & \multicolumn{1}{c|}{5.181}     & \multicolumn{1}{c|}{5.465}     & 7.018     \\ \hline
\end{tabular}
\label{tab:table-vi}
\end{table}

Prioritising scientific validity, we therefore selected the (\texttt{Spec-Depth=4}, \texttt{Spec-Width=256}) configuration as the robust foundation. The fine-tuning process was completed with a coordinate ascent tuning of the Photometric and Fusion module complexities, which identified a 2-layer MLP and a 4-layer Fusion module as the optimal configurations. This multi-stage optimisation process yielded the final model architecture for Phase~III, defined by the (\texttt{Spec-Depth=4}, \texttt{Spec-Width=256}, \texttt{MLP-Depth=2}, \texttt{MLP-Width=256}, \texttt{Fusion-Depth=4}, \texttt{Fusion-Width=512}) parameter configuration.

\subsection{Phase III: Final Model Performance and the Precision-Realism Trade-Off}
The final phase of this research transitioned to the full, $166{,}625$-star dataset, aiming to train the optimised best-performing architecture to its maximum potential. As suggested in our methodology, this phase involved an exploratory and iterative process to identify a stable and effective multi-task training strategy capable of handling the severe, real-world class imbalance.

Our initial explorations focused on advanced training strategies, including a hybrid balancing approach, which applies the principles of transfer learning \citep{howard2018universal} by pre-training on a source dataset before fine-tuning. We also explored various staged-learning methodologies, inspired by the concept of Curriculum Learning \citep{bengio2009curriculum}, where the model is exposed to progressively more complex tasks. However, these attempts consistently proved to be unstable, often resulting in a catastrophic degradation of the model's regression capabilities upon the re-introduction of the multi-task objective. A refined fine-tuning of our best end-to-end model from Phase~I (Model~6) using a differential learning rate scheme also yielded only marginal improvements. 

These preliminary results led us to a critical insight: the primary bottleneck appeared not to be the training strategy itself (e.g., staged vs. end-to-end), but rather the intrinsic conflict between the regression and classification objectives, governed by the composition of the multi-task loss function.

Based on this insight, we hypothesised that the relative weighting of the task-specific losses was the most critical factor. To systematically investigate this, we conducted a final controlled experiment, comparing three models with different multi-task loss weightings: Setup~A (regression-dominant, $0.5{:}0.4{:}0.1$), Setup~B (moderately balanced, $0.4{:}0.4{:}0.2$), and Setup~C (most balanced, $0.35{:}0.35{:}0.3$).

\begin{figure}[t]
    \centering
    \includegraphics[width=\linewidth]{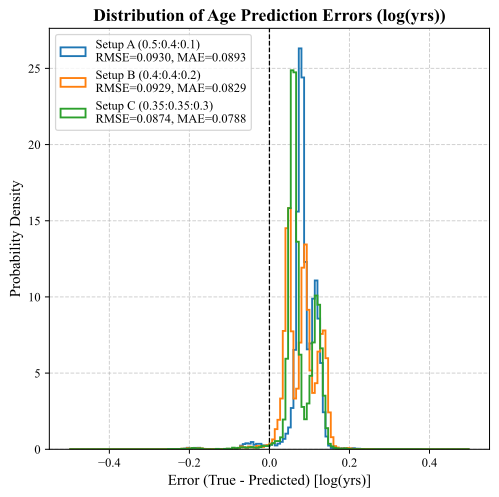}
    \caption{Distribution of age prediction errors in the logarithmic space ($\log(\mathrm{yrs})$) for the final three model configurations.}
    \label{fig:age_errors}
\end{figure}

As shown in Fig.~\ref{fig:age_errors} and Table~\ref{tab:table-vii}, a clear trade-off emerged. Setup~C achieved the best mathematical fitting performance in the $\log(\mathrm{yrs})$ space, attaining the lowest RMSE and MAE. However, this superior mathematical precision came at a fatal cost: a complete loss of physical realism, with predicted maximum ages exceeding $2700~\mathrm{Gyr}$.

\begin{table}[t]
\centering
\caption{Comprehensive performance metrics for Setup A, B, and C.}
\begin{tabular}{|llccc|}
\hline
\multicolumn{2}{|c|}{Metrics}                                                                                                            & \multicolumn{1}{c|}{Setup A} & \multicolumn{1}{c|}{Setup B}  & Setup C  \\ \hline
\multicolumn{2}{|l|}{Age RMSE (log(yrs))}                                                                                                & \multicolumn{1}{c|}{0.0930}  & \multicolumn{1}{c|}{0.0929}   & 0.0874   \\ \hline
\multicolumn{2}{|l|}{Lifetime RMSE (log(yrs))}                                                                                           & \multicolumn{1}{c|}{0.0973}  & \multicolumn{1}{c|}{0.0953}   & 0.0893   \\ \hline
\multicolumn{2}{|l|}{Classification F1 (macro)}                                                                                          & \multicolumn{1}{c|}{0.4998}  & \multicolumn{1}{c|}{0.5927}   & 0.6157   \\ \hline
\multicolumn{5}{|c|}{True Age Range (Gyr): 2.842 $\sim$6.107; Mean: 5.022}                                                                                                                                         \\ \hline
\multicolumn{1}{|l|}{\multirow{3}{*}{\begin{tabular}[c]{@{}l@{}}Predicted\\ Age Range\\ (Gyr)\end{tabular}}} & \multicolumn{1}{l|}{Min}  & \multicolumn{1}{c|}{2.320}   & \multicolumn{1}{c|}{1.674}    & 2.675    \\ \cline{2-5} 
\multicolumn{1}{|l|}{}                                                                                       & \multicolumn{1}{l|}{Max}  & \multicolumn{1}{c|}{5.703}   & \multicolumn{1}{c|}{1419.738} & 2721.162 \\ \cline{2-5} 
\multicolumn{1}{|l|}{}                                                                                       & \multicolumn{1}{l|}{Mean} & \multicolumn{1}{c|}{4.171}   & \multicolumn{1}{c|}{4.321}    & 4.436    \\ \hline
\end{tabular}
\label{tab:table-vii}
\end{table}

In stark contrast, Setup~A was the only configuration to produce predictions that remained within a scientifically valid range. While its $\log(\mathrm{yrs})$ RMSE was marginally higher and its F1-score ($0.4998$) was the lowest of the three, its foundation in physical realism makes it the only scientifically viable candidate. This critical finding underscores that for scientific applications, a model’s adherence to physical constraints must be a primary evaluation criterion. Based on its scientific validity and robust regression performance, Setup~A was definitively selected as the best-performing.

\subsection{Uncertainty Quantification}
While the performance metrics presented in the previous section quantify the model~6's overall predictive accuracy, a critical aspect for any scientific machine learning system is its ability to convey the reliability of its individual predictions. To this address this, we implemented an uncertainty quantification (UQ) analysis to assess the predictive confidence of our final best-performing model. This process is important for establishing the model's reliability and providing users with an approach to identify potentially unreliable estimates.

The adopted methodology was Monte Carlo (MC) Dropout, a technique that leverages the stochastic nature of dropout layers during inference to approximate the posterior distribution of the model's predictions \citep{gal2016dropout}. For each sample in the test set, we performed 50 forward passes with dropout enabled, generating a distribution of predictions for each of the three output tasks. The mean of this distribution was taken as the final prediction, while the standard deviation was used as a measure of the model's knowledge uncertainty --- a reflection of the model's own confidence in its prediction based on the training data it has seen.

The results of this analysis indicate that our model is well-calibrated and that its uncertainty estimates are scientifically meaningful.

\begin{figure*}[t]
    \centering
    \begin{subfigure}{0.48\textwidth}
        \centering
        \includegraphics[width=\linewidth]{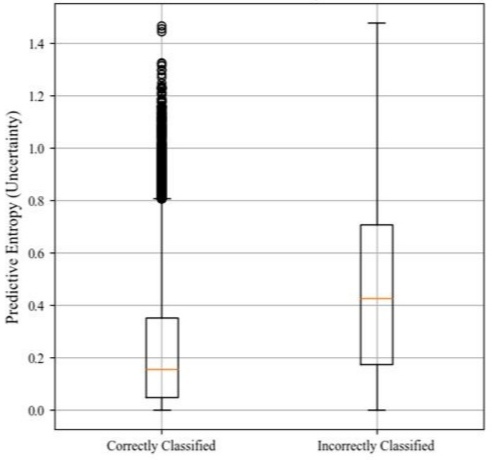}
        \caption{Distribution of classification uncertainty measured by predictive entropy.}
        \label{fig:uncertainty_a}
    \end{subfigure}
    \hfill
    \begin{subfigure}{0.48\textwidth}
        \centering        \includegraphics[width=\linewidth]{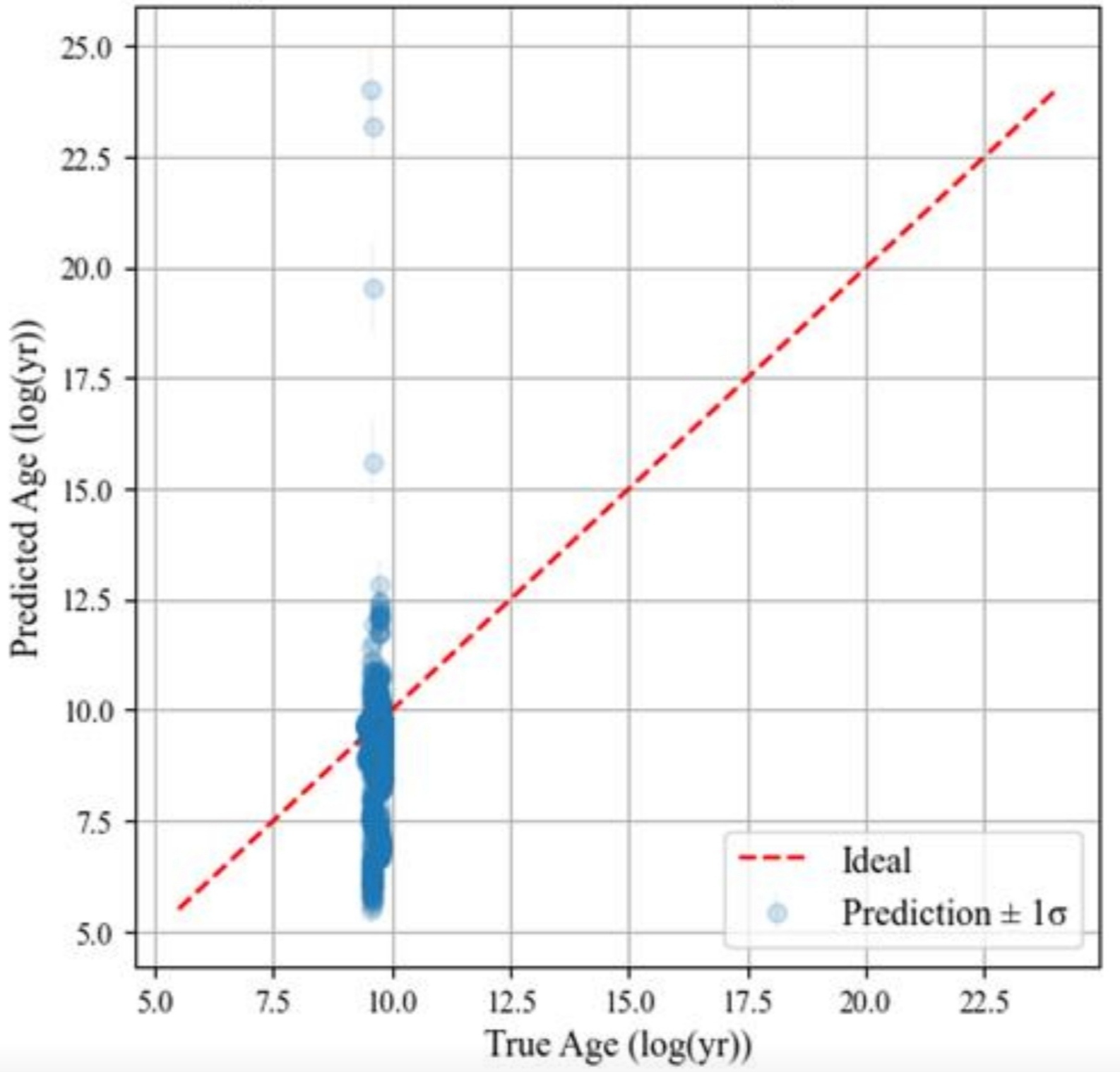}
        \caption{Age regression performance with quantified uncertainty on the test set.}
        \label{fig:uncertainty_b}
    \end{subfigure}
    \caption{Uncertainty quantification results for the final best-performing model (Setup~A).}
    \label{fig:uncertainty_combined}
\end{figure*}

\begin{enumerate}

    \item Classification Uncertainty: The analysis of classification uncertainty reveals that the model expresses higher uncertainty when it makes incorrect predictions. As shown in Fig.~\ref{fig:uncertainty_a}, we computed the predictive entropy --- a measure of the sharpness of the predicted probability distribution --- for all test samples. The resulting box plot clearly shows that the distribution of uncertainty for incorrectly classified samples is shifted towards higher entropy values than for correctly classified ones. This indicates that the model tends to be less confident when it is making a mistake, and this providing a valuable indicator of prediction reliability.
    
    \item Regression Uncertainty: For the regression tasks, the UQ analysis provides a per-prediction credible interval, this offering an important evaluation metric for each individual predicted stellar age and lifetime. Fig.~\ref{fig:uncertainty_b} shows the age prediction scatter plot in $\log(\mathrm{yrs})$ space, now augmented with error bars that representing one standard deviation ($(\pm 1\sigma)$) of the MC Dropout predictions. We observed that while most predictions show high confidence (small error bars), a subset of samples shows significantly larger uncertainty. These high-uncertainty predictions typically correspond to the stars with more ambiguous features or those located in the zones of the parameter space that are sparsely represented in the training data.

\end{enumerate}

The ability to highlight which predictions are more or less reliable transforms our proposed model from a simple predictor into a scientifically useful tool. It enables researchers to focus on the most reliable parameter estimations, hence making it an important requirement for robust scientific inquiry.

\section{Conclusions and Future Work}
This project adopted a systematic three-phase approach to develop and evaluate a deep multimodal neural network for the joint estimation of stellar age, lifetime, and evolutionary stage from SDSS data. Our work navigated the complex relationship between data balancing strategies, architectural complexity, and the intrinsic conflicts of multi-task learning. The research led to the identification of a scientifically robust final model, whose development and evaluation provide several key conclusions and recommendations for future work.

The primary outcomes and contributions of this research can be synthesised into three main findings. First, our empirical results demonstrated that for this dataset, a data-level under sampling strategy unexpectedly outperformed an algorithm-level weighted-loss approach, highlighting that a bias-free learning environment can be more critical than retaining all data under conditions of severe class imbalance. Second, the systematic fine-tuning process revealed the necessity of a comprehensive evaluation criterion; it was discovered that optimising for standard mathematical metrics alone can lead to models that are physically unrealistic, establishing scientific validity as an important principle for model selection. Finally, our work identified a fundamental trade-off between regression precision and classification accuracy in a unified multi-task framework, suggesting an intrinsic conflict in the feature representations required for these different task types.

Based on these findings, our best-performing model was deliberately selected for its optimal balance between regression accuracy and physical reliability. It achieved a state-of-the-art performance on its primary regression tasks, with an Age RMSE of $0.093$ in $\log(\mathrm{yrs})$ space. Its adherence to astrophysical constraints, coupled with quantified uncertainty via Monte Carlo Dropout, makes it as a scientifically robust and useful tool. While its classification capability (Macro $\mathrm{F1} \approx 0.5$) remains a notable limitation, this reflects the intrinsic ambiguity of the stage labels rather than a model deficiency.

Looking forward, the observed tension between tasks strongly motivates future work into decoupled or expert model architectures, where classification and regression might be handled by specialised models. Or, explore shared-but-constrained representations (e.g., through structured representation learning). Furthermore, the integration of complementary data from other surveys, such as Gaia and APOGEE, promises to further constrain the problem and enhance model performance. In conclusion, this work not only delivers a scientifically robust framework for stellar parameter estimation but also provides critical, empirically-grounded insights into the methodological challenges and evaluation strategies required for applying deep learning to complex, real-world scientific problems.

\bibliographystyle{aasjournal}
\bibliography{references}

@book{maeder2008physics,
  title={Physics, formation and evolution of rotating stars},
  author={Maeder, Andr{\'e}},
  year={2008},
  publisher={Springer Science \& Business Media}
}

@article{soderblom2010ages,
  title={The ages of stars},
  author={Soderblom, David R},
  journal={Annual Review of Astronomy and Astrophysics},
  volume={48},
  number={1},
  pages={581--629},
  year={2010},
  publisher={Annual Reviews}
}

@article{ksoll2020stellar,
  title={Stellar parameter determination from photometry using invertible neural networks},
  author={Ksoll, Victor F and Ardizzone, Lynton and Klessen, Ralf and Koethe, Ullrich and Sabbi, Elena and Robberto, Massimo and Gouliermis, Dimitrios and Rother, Carsten and Zeidler, Peter and Gennaro, Mario},
  journal={Monthly Notices of the Royal Astronomical Society},
  volume={499},
  number={4},
  pages={5447--5485},
  year={2020},
  publisher={Oxford University Press}
}

@article{gai2024simultaneous,
  title={Simultaneous derivation of galaxy physical properties with multimodal deep learning},
  author={Gai, Mario and Bove, Mario and Bonetta, Giovanni and Zago, Davide and Cancelliere, Rossella},
  journal={Monthly Notices of the Royal Astronomical Society},
  volume={532},
  number={2},
  pages={1391--1401},
  year={2024},
  publisher={Oxford University Press}
}

@article{stoughton2002sloan,
  title={Sloan digital sky survey: early data release},
  author={Stoughton, Chris and Lupton, Robert H and Bernardi, Mariangela and Blanton, Michael R and Burles, Scott and Castander, Francisco J and Connolly, AJ and Eisenstein, Daniel J and Frieman, Joshua A and Hennessy, GS and others},
  journal={The Astronomical Journal},
  volume={123},
  number={1},
  pages={485},
  year={2002},
  publisher={IOP Publishing}
}

@article{li2025machine,
  title={Machine Learning in Stellar Astronomy: Progress up to 2024},
  author={Li, Guangping and Lu, Zujia and Wang, Junzhi and Wang, Zhao},
  journal={arXiv preprint arXiv:2502.15300},
  year={2025}
}

@article{bell2001stellar,
  title={Stellar Mass-to-Light Ratios and theTully-Fisher Relation},
  author={Bell, Eric F and de Jong, Roelof S},
  journal={The Astrophysical Journal},
  volume={550},
  number={1},
  pages={212},
  year={2001},
  publisher={IOP Publishing}
}

@article{chaplin2013asteroseismology,
  title={Asteroseismology of solar-type and red-giant stars},
  author={Chaplin, William J and Miglio, Andrea},
  journal={Annual Review of Astronomy and Astrophysics},
  volume={51},
  pages={353--392},
  year={2013},
  publisher={Annual Reviews}
}

@article{barnes2008gyrochronology,
  title={Gyrochronology and its usage for main sequence field star ages},
  author={Barnes, Sydney A},
  journal={Proceedings of the International Astronomical Union},
  volume={4},
  number={S258},
  pages={345--356},
  year={2008},
  publisher={Cambridge University Press}
}

@article{lecun2015deep,
  title={Deep learning},
  author={LeCun, Yann and Bengio, Yoshua and Hinton, Geoffrey},
  journal={nature},
  volume={521},
  number={7553},
  pages={436--444},
  year={2015},
  publisher={Nature Publishing Group UK London}
}

@article{wang2024deep,
  title={Deep learning prediction of galaxy stellar populations in the low-redshift Universe},
  author={Wang, Li-Li and Yang, Guang-Jun and Zhang, Jun-Liang and Rong, Li-Xia and Zheng, Wen-Yan and Liu, Cong and Chen, Zong-Yi},
  journal={Monthly Notices of the Royal Astronomical Society},
  volume={527},
  number={4},
  pages={10557--10563},
  year={2024},
  publisher={Oxford University Press}
}

@inproceedings{he2016deep,
  title={Deep residual learning for image recognition},
  author={He, Kaiming and Zhang, Xiangyu and Ren, Shaoqing and Sun, Jian},
  booktitle={Proceedings of the IEEE conference on computer vision and pattern recognition},
  pages={770--778},
  year={2016}
}

@article{vaswani2017attention,
  title={Attention is all you need},
  author={Vaswani, Ashish and Shazeer, Noam and Parmar, Niki and Uszkoreit, Jakob and Jones, Llion and Gomez, Aidan N and Kaiser, {\L}ukasz and Polosukhin, Illia},
  journal={Advances in neural information processing systems},
  volume={30},
  year={2017}
}

@article{caruana1997multitask,
  title={Multitask learning},
  author={Caruana, Rich},
  journal={Machine learning},
  volume={28},
  number={1},
  pages={41--75},
  year={1997},
  publisher={Springer}
}

@article{choi2016mesa,
  title={Mesa isochrones and stellar tracks (MIST). I. Solar-scaled models},
  author={Choi, Jieun and Dotter, Aaron and Conroy, Charlie and Cantiello, Matteo and Paxton, Bill and Johnson, Benjamin D},
  journal={The Astrophysical Journal},
  volume={823},
  number={2},
  pages={102},
  year={2016},
  publisher={IOP Publishing}
}

@article{savitzky1964smoothing,
  title={Smoothing and differentiation of data by simplified least squares procedures.},
  author={Savitzky, Abraham and Golay, Marcel JE},
  journal={Analytical chemistry},
  volume={36},
  number={8},
  pages={1627--1639},
  year={1964},
  publisher={ACS Publications}
}

@inproceedings{dauphin2017language,
  title={Language modeling with gated convolutional networks},
  author={Dauphin, Yann N and Fan, Angela and Auli, Michael and Grangier, David},
  booktitle={International conference on machine learning},
  pages={933--941},
  year={2017},
  organization={PMLR}
}

@article{loshchilov2017decoupled,
  title={Decoupled weight decay regularization},
  author={Loshchilov, Ilya and Hutter, Frank},
  journal={arXiv preprint arXiv:1711.05101},
  year={2017}
}

@book{geron2022hands,
  title={Hands-on machine learning with Scikit-Learn, Keras, and TensorFlow},
  author={G{\'e}ron, Aur{\'e}lien},
  year={2022},
  publisher={" O'Reilly Media, Inc."}
}

@inproceedings{gal2016dropout,
  title={Dropout as a bayesian approximation: Representing model uncertainty in deep learning},
  author={Gal, Yarin and Ghahramani, Zoubin},
  booktitle={international conference on machine learning},
  pages={1050--1059},
  year={2016},
  organization={PMLR}
}

@article{howard2018universal,
  title={Universal language model fine-tuning for text classification},
  author={Howard, Jeremy and Ruder, Sebastian},
  journal={arXiv preprint arXiv:1801.06146},
  year={2018}
}

@inproceedings{bengio2009curriculum,
  title={Curriculum learning},
  author={Bengio, Yoshua and Louradour, J{\'e}r{\^o}me and Collobert, Ronan and Weston, Jason},
  booktitle={Proceedings of the 26th annual international conference on machine learning},
  pages={41--48},
  year={2009}
}

\end{document}